\documentstyle[12pt]{article}
\renewcommand\theequation{\thesection.\arabic{equation}}

\begin{document}
\setlength{\unitlength}{1mm}

{\hfill   April 1996 }

{\hfill    WATPHYS-TH-96-04 }

{\hfill    hep-th/9604118} \vspace*{2cm} \\
\begin{center}
{\Large\bf
Conical geometry and quantum entropy of a charged Kerr black hole}
\end{center}
\begin{center}
{\large\bf  Robert B.~Mann\footnote{e-mail: mann@avatar.uwaterloo.ca} and Sergey N.~Solodukhin\footnote{e-mail: sergey@avatar.uwaterloo.ca}\footnote{NATO Fellow}}
\end{center}
\begin{center}
{\it Department of Physics, University of Waterloo, Waterloo, Ontario N2L 3G1, 
Canada}  

\medskip

\end{center}
\vspace*{2cm}
\begin{abstract}
We apply the method of conical singularities to calculate the 
tree-level entropy and
its one-loop quantum corrections for a charged Kerr black hole.
The Euclidean geometry for the Kerr-Newman metric is considered. 
We show that for an arbitrary periodization in Euclidean space there
exists a conical singularity at the
horizon. Its $\delta$-function like curvatures are calculated and are shown to 
behave similar to the static case. The heat kernel expansion for a scalar field on this
conical space background is derived and the (divergent) quantum correction
to the entropy is obtained. It is argued that 
these divergences can be removed by renormalization of couplings 
in the tree-level gravitational action in a manner similar to that 
for a static black hole. 
\end{abstract}
\begin{center}
{\it PACS number(s): 04.70.Dy, 04.62.+v}
\end{center}
\vskip 1cm
\newpage
\baselineskip=.8cm
\section{Introduction}
\setcounter{equation}0

The notion that  black holes could be considered as a thermodynamic systems
characterized by temperature, energy and entropy was first proposed by 
Bekenstein \cite{1} and confirmed via the discovery of their
thermal radiation properties by Hawking \cite{2}. Independently, it was 
realized that there are only a few macroscopic parameters
which can be assigned to a black hole: its mass ($m$), charge ($q$)
and angular momentum ($\Omega$). In the static case, angular momentum 
vanishes. A typical representative of this class is the Reissner-Nordstrom 
black hole which is a solution of the Einstein equations with a 
Maxwell field as a source.
Such a hole is characterized by just its mass ($m$) and charge ($q$).

When rotation is present, the Einstein-Maxwell equations have
the Kerr-Newman metric \cite{KN} as a solution. This metric
corresponds to a black hole of a general type characterized by all three 
parameters ($m,q,\Omega$).
Remarkably, the thermodynamic analogy works for this general case;
in particular, it suggests that there is an entropy associated with 
this hole that is proportional to the area of the horizon. 
If this analogy is exact, there must be hidden degrees of freedom 
of the  hole which are counted by the Bekenstein-Hawking entropy.
Recently, there has been much interest in attempting to provide a 
statistical explanation of these degrees of freedom and their relationship
to the entropy (see reviews \cite{4}, \cite{5}) within some 
quantum-mechanical calculations \cite{6}-\cite{FS}. However, the proposed 
expressions for the entropy can be considered to be quantum (one-loop) 
corrections to the classical quantity, and do not give any explanation of
the classical entropy itself.

According to 't Hooft \cite{6}, one can relate the entropy of a black hole 
with a thermal gas of quantum field excitations propagating outside the 
horizon. In his model 't Hooft introduced a ``brick wall'' cut-off: a fixed 
boundary near the horizon within which the quantum field does not propagate. 
Its role is to eliminate divergences which appear due to the
infinite growth of the density of states close to the horizon.
This model can be successfully formulated in different space-time dimensions
\cite{7}. The quantization of a field system typically requires an 
ultraviolet (UV) regularization procedure that must be taken into account in the 
statistical-mechanical calculation as well. Remarkably, it was demonstrated 
in \cite{14} that the Pauli-Villars regularization not only removes the 
standard field-theoretical UV-divergences but automatically implements a 
cut-off in the 't Hooft calculation, rendering  unnecessary
the introduction of the ``brick wall''.

The natural way to formulate black hole thermodynamics is to use the
Euclidean path integral approach \cite{GH}. For an arbitrary field system
it entails  closing the Euclidean time coordinate with a period $\beta=T^{-1}$
where $T$ is the temperature of the system. This yields a periodicity
condition for the quantum fields in the path integral. In the black hole 
case for arbitrary $\beta$ this procedure leads to an effective
Euclidean manifold which has a conical singularity at the horizon that 
vanishes for a fixed value $\beta=\beta_H$. Thermodynamic quantities 
({\it i.e.} energy and entropy)
are calculated by differentiating the corresponding free energy with 
respect to $\beta$ and then setting $\beta=\beta_H$. This procedure was 
consistently carried out for a static black hole and resulted in obtaining 
the general UV-divergent structure of the entropy \cite{SS}, \cite{F}, 
\cite{SS1}, \cite{FS}.

Essentially, the divergences of the entropy have the same origin as the
UV-divergences of the quantum effective action and can be removed by 
remormalization of the gravitational couplings in the tree-level 
gravitational action \cite{9}, \cite{J}, \cite{FS}. The technique 
developed in \cite{FS} allowed proof of this statement
for an arbitrary static black hole. Alternatively, this was demonstrated 
for the Reissner-Nordstrom black hole within 't Hooft's approach \cite{14} 
applying the Pauli-Villars regularization scheme.

An essential loophole in the above considerations is that 
they were concerned with only static, non-rotating black holes.
The only exception is a series of recent preprints \cite{K} where
't Hooft's approach was applied to a Kerr-Newman black hole and some 
(qualitative) analysis of divergences was presented.
Adoption of conical methods for stationary black holes necessitates 
dealing with problems of treating Euclideanization 
(or complexification) of the Kerr-Newman metric \cite{E} and a
general periodicity analysis of its conical geometry. Although the passage 
to a Euclidean metric and periodicity arguments were given some time ago
\cite{GH} the conical geometry for arbitrary period in the Kerr-Newman case 
remains unclear. Additional outstanding questions include
the structure of the UV-divergences of the entropy of stationary black holes
and whether or not their renormalization works in the same way as 
for a static hole.

In this paper we consider these questions in detail. 
In Section 2 we describe the passage to Euclidean 
space for the Kerr-Newman metric, establishing the structure of this space 
in the vicinity of the horizon. We determine the conditions necessary for
periodicity in the direction of the 
time-like Killing vector (which is analog of
the Euclidean time vector $\partial_\tau$ in the static case) 
corresponding to regular Euclidean space. For arbitrary periods there 
is a conical singularity at the horizon surface.  The geometry of this 
conical space is studied in Section 3. We employ the regularization method  
suggested in \cite{FS1} and obtain the expected
$\delta$-function like behavior 
of the curvatures. The integrals of quadratic combinations of curvature 
tensors are also considered. The results we obtain have a marked similarity 
to the static case. In Section 4 we consider the Euclidean path integral 
quantization of a scalar matter field in the background of a 
conical Kerr-Newman metric. We obtain the UV-divergences for the entropy, 
the structure of which is similar to that obtained in the static case. 
We argue that  these divergences of entropy
are renormalized in the same way as 
for a static black hole.

\section{Euclidean Kerr-Newman geometry}
\setcounter{equation}0

The Kerr-Newman metric of the space-time with Minkowskian signature in 
Boyer-Lindquist coordinates takes the form:
\begin{eqnarray}
&&ds^2=g_{rr}dr^2+g_{\theta\theta}d\theta^2 + g_{tt}dt^2+2g_{t\phi}dtd\phi +g_{\phi\phi}d\phi^2
\nonumber \\
&&g_{rr}={\rho^2 \over \Delta}~~,~~ g_{\theta\theta}=\rho^2~~,~~g_{tt}=-{( \Delta-
a^2\sin^2\theta ) \over  \rho^2}~~; \nonumber \\
&&g_{t\phi}=-{a \sin^2\theta (r^2+a^2-\Delta ) \over \rho^2 }~, ~~g_{\phi\phi}=\left(
{(r^2+a^2)^2-\Delta a^2\sin^2 \theta
\over \rho^2}\right) \sin^2 \theta~, \nonumber \\
&&\Delta (r)=r^2+a^2+q^2-2mr~, ~~\rho^2=r^2+a^2 \cos^2 \theta
\label{1}
\end{eqnarray}
The function $\Delta (r)$ can be represented in the form $\Delta (r)=(r-r_+)(r-r_-)$,
where $r_{\pm}=m\pm \sqrt{m^2-a^2-q^2}$.

This space-time has a pair of orthogonal Killing vectors:
\begin{eqnarray}
&&K=\partial_t+{a\over r^2+a^2}\partial_\phi~,~~\tilde{K}=a\sin^2\theta \partial_t+\partial_\phi
 \nonumber \\
&&K^2=-{\Delta \rho^2 \over (r^2+a^2)^2}~,
~~\tilde{K}^2=\rho^2\sin^2\theta~, ~~K\cdot\tilde{K}=0~~.
\label{2}
\end{eqnarray}
The vector $K$ is time-like everywhere in the region $r\geq r_+$ and 
becomes null $K^2=0$ for $r=r_+$, whereas
$\tilde{K}$ is space-like everywhere outside the 
axis ($\theta=0,~\theta=\pi$) where $\tilde{K}^2=0$.
The one-forms dual to $K$ and  $\tilde{K}$ are
\begin{eqnarray}
&&\omega={(r^2+a^2)\over \rho^2}(dt-a\sin^2\theta d \phi ) \nonumber \\
&&\tilde{\omega}={(r^2+a^2)\over \rho^2}(d\phi-{a\over (r^2+a^2)}dt) \nonumber \\
&&\omega [K]=\tilde{\omega}[\tilde{K}]=1~,~~\omega [\tilde{K}]=\tilde{\omega}[K]=0
\label{3}
\end{eqnarray}

To obtain the correspondence with the Schwarzschild metric 
note that $K$ and $\tilde{K}$ are the respective analogs of the
vectors $\partial_t$ and $\partial_\phi$ and
$\omega$ and $\tilde{\omega}$ are the respective analogs 
of $dt$ and $d\phi$ of the Schwarzschild metric.
This correspondence is almost exact with one exception: 
$\omega$ and $\tilde{\omega}$ together with $d\theta$ and $dr$ form 
an anholonomic basis of one-forms. This means that there are
no globally defined coordinates $X$ and $\tilde{X}$ such that 
$\omega=dX$ and $\tilde{\omega}=d\tilde{X}$.

Horizon surfaces $\Sigma$ are defined as the surfaces where the 
time-like vector $K$ becomes null, $K^2|_\Sigma=0$; the outer horizon 
is the surface for which $r=r_+$. In addition to this one often considers
the surface where the vector
$\partial_t$ becomes null. This surface is called the ergosphere and is 
determined by equation $r^2+a^2\cos^2\theta+q^2-2mr=0$. It lies outside 
the outer horizon $\Sigma$, touching it at the axis $\theta=0$ and 
$\theta=\pi$.


Consider now the Euclideanization of the Kerr-Newman metric. The standard 
prescription says \cite{GH} that we must change the time variable $t=\imath \tau$ 
and supplement this by the parameter transformation 
$a=\imath \hat{a},~q=\imath \hat{q}$.
The {\it Euclidean} vectors $K,~\tilde{K}$ and the 
corresponding one-forms $\omega, ~~\tilde{\omega}$ take the form:
\begin{eqnarray}
&&K=\partial_\tau-{\hat{a}\over r^2-\hat{a}^2}\partial_\phi~,~~\tilde{K}=\hat{a}\sin^2\theta 
\partial_\tau+\partial_\phi \nonumber \\ 
&&\omega={(r^2-\hat{a}^2)\over \hat{\rho}^2}(d\tau-\hat{a}\sin^2\theta d \phi ) \nonumber \\
&&\tilde{\omega}={(r^2-\hat{a}^2)\over \hat{\rho}^2}(d\phi+{\hat{a}\over (r^2-\hat{a}^2)}d\tau) 
\label{4}
\end{eqnarray}
where $\hat{\rho}^2=r^2-\hat{a}^2\cos^2\theta$.
The Euclidean metric can be written in the form:
\begin{equation}
ds^2_E={\hat{\rho}^2\over \hat{\Delta}}dr^2+{\hat{\Delta}\hat{\rho}^2\over
(r^2-\hat{a}^2)^2}\omega^2+\hat{\rho}^2(d\theta^2+\sin^2\theta \tilde{\omega}^2)
\label{5}
\end{equation}
where $\omega$ and $\tilde{\omega}$ take the form (\ref{4}), $\hat{\Delta}=r^2-\hat{a}^2
-\hat{q}^2-2mr$.
Roots of the function $\hat{\Delta}$ are now 
$\hat{r}_\pm=m\pm\sqrt{m^2+\hat{a}^2+\hat{q}^2}$.
The horizon surface $\Sigma$ defined by $r=\hat{r}_+$ is the stationary 
surface of the Killing vector $K$. Consider  metric (\ref{5}) near 
$r=\hat{r}_+$. It is useful to introduce a new {\it radial}
variable $x$ such that near the horizon we have
\begin{eqnarray}
&&\hat{\Delta}=\gamma (r-\hat{r}_+)={\gamma^2 x^2 \over 4}, \nonumber \\
&&(r-\hat{r}_+)={\gamma x^2 \over 4},~~\gamma=2\sqrt{m^2+\hat{a}^2+\hat{q}^2}
\label{6}
\end{eqnarray}
Then the metric (\ref{5}) up to terms $O(x^2)$ reads:
\begin{equation}
ds^2_E=ds^2_\Sigma+\hat{\rho}^2_+ \left( dx^2+{\gamma^2 x^2 \over 4(\hat{r}^2_+-\hat{a}^2)^2}
\omega^2 \right) ~~,
\label{7}
\end{equation}
where $\hat{\rho}^2_+=\hat{r}^2_+-\hat{a}^2\cos^2\theta$ and
\begin{eqnarray}
&&ds^2_\Sigma=\hat{\rho}_+^2 \left(d\theta^2+\sin^2\theta \tilde{\omega}^2 \right) \nonumber \\
&&=\hat{\rho}_+^2d\theta^2+{(\hat{r}^2_+-\hat{a}^2)^2\over\hat{\rho}_+^2}   \sin^2\theta
d\psi^2
\label{8}
\end{eqnarray}
is metric on the horizon surface $\Sigma$. In writing (\ref{8}) 
we employed the fact that on $\Sigma$ we may
introduce the well-defined angle coordinate 
$\psi=\phi+{\hat{a} \over (\hat{r}^2_+-\hat{a}^2)} \tau$.
The regularity of the metric (\ref{8}) at the points 
$\theta=0,~~\theta=\pi$ requires the identification of the points 
$\psi$ and $\psi +2\pi$ on $\Sigma$.
After all calculations with the Euclideanized Kerr-Newman have been 
completed, we analytically continue the results obtained 
back to the real values $a$ and $q$.

Expression (\ref{7}) may be rewritten as follows:
\begin{equation}
ds^2_E=ds^2_\Sigma+\hat{\rho}^2_+ds^2_{C_2}
\label{9}
\end{equation}
where $ds^2_{C_2}$ is metric of a two-dimensional disk $C_2$ attached to the horizon $\Sigma$ at a point ($\theta,~\psi$):
\begin{equation}
ds^2_{C_2}=dx^2+{\gamma^2 x^2\over 4 \hat{\rho}_+^4}(d\tau-\hat{a}\sin^2\theta d\phi )^2~~.
\label{10}
\end{equation}
Confider the metric (\ref{10}) with ($\theta,~\psi )$ fixed. 
Then we may introduce an
angle coordinate on $C_2$, $\chi=\tau-\hat{a}\sin^2\theta\ \phi$, in terms of which the metric 
reads
\begin{equation}
ds^2_{C_2}=dx^2+{\gamma^2 x^2\over 4 \hat{\rho}_+^4}d\chi^2~~.
\label{11}
\end{equation}
Requiring the absence of a conical singularity at $x=0$ means that we 
must identify points
$\chi$ and $\chi+4\pi\gamma^{-1}\hat{\rho}^2_+$. In order for this to 
hold independently of the
coordinate $\theta$ on the horizon we must also
identify points $(\tau,~\phi )$
with $(\tau+2\pi \beta_H,~\phi-2\pi \Omega \beta_H)$, where $\Omega={\hat{a} \over(
\hat{r}^2_+- \hat{a}^2)}$ is the
(complex) angular velocity and $\beta_H={(\hat{r}_+^2-\hat{a}^2) \over 
\sqrt{m^2+\hat{a}^2+\hat{q}^2}}$. It is easy to see that the identified points have the same coordinate $\psi$.

With the described identification we obtain the following picture of the Euclidean
 Kerr-Newman geometry
in the vicinity of the horizon $\Sigma$. Attached to every point $(\theta, \psi$) 
of the horizon is a two-dimensional disk $C_2$ with coordinates
($x, \chi$). Although $\chi$ is not a global coordinate in four-dimensional 
space  and at each point $(\theta, \psi )$
there is a new $\chi$, the periodic identification of points on $C_2$ 
works universally and independently of any point on the horizon $\Sigma$.
As in static case, the Euclidean Kerr-Newman geometry possesses an abelian 
isometry generated by the Killing vector $K$ with 
horizon surface $\Sigma$ being a fixed set of the isometry. 
Globally, $K$ is not a coordinate vector. However,
locally we have $K=\partial_\chi$ where $\chi$ was introduced above.
The periodicity is in the direction of the vector $K$ and
the resulting Euclidean space $E$ is regular manifold.

\bigskip

\section{Conical singularity and curvature tensors} 
\setcounter{equation}0

Assume now that we close the trajectory of the Killing vector $K$ with
arbitrary period $2\pi\beta$.
Near the horizon this means that on $C_2$ in (\ref{9}), (\ref{10}) 
we identify points $(\tau,~\phi )$ and $(\tau+2\pi \beta,~\phi-2\pi
 \Omega \beta)$ with $\beta \neq \beta_H$. Note again that 
points identified in this way
have the same value of the coordinate $\psi$. Then $\chi$ is 
an angle coordinate
with period $2\pi\beta (1+\hat{a}\Omega \sin^2\theta )$. By introducing a
new angle coordinate
$\chi=\beta \hat{\rho}^2_+ (\hat{r}_+^2-\hat{a}^2)^{-1}\bar{\chi}$ 
which has period $2\pi$, the metric on $C_2$ becomes
\begin{equation}
ds^2_{C_{2,\alpha}}=dx^2+\alpha^2x^2d\bar{\chi}^2
\label{12}
\end{equation}
and coincides with the metric on a two dimensional cone with 
angular deficit $\delta=2\pi (1-\alpha)$, $\alpha={\beta\over \beta_H}$.
With the above identification the four-dimensional metric (\ref{5}) 
describes the Euclidean conical space $E_\alpha$ with singular 
surface $\Sigma$.

Curvature tensors at conical singularities behave as distributions. 
This behavior was precisely established for a flat 2d cone in 
\cite{Sokolov-Starobinski} and for a general static metric in \cite{FS1}. 
The Kerr-Newman metric, which is the subject of
our consideration here, is stationary and not static. Therefore, all the 
formulae obtained in \cite{FS1} must be checked for this case. 

To proceed, we apply the method which was
successful in the static case (see details in \cite{FS1}). 
It consists in regulating the conical singularity when the 
cone metric (\ref{12}) is replaced by a sequence of regular metrics
labeled by a parameter $b$:
\begin{equation}
ds^2_{C_{2,\alpha , b}}=f(x,b)dx^2+\alpha^2x^2d\bar{\chi}^2~~,
\label{13}
\end{equation}
where $f(x,b)$ is some smooth regulating function that approaches
unity as $b\to 0$, {\it e.g.}
\begin{equation}
f(x,b)={x^2+\alpha^2 b^2 \over x^2+b^2}
\label{14}
\end{equation}
is a suitable regularization function.
In the limit $b\rightarrow 0$ the sequence of metrics (\ref{13}) 
re-produces a $\delta$-function-like contribution
to the curvature.

Applying this method to the Kerr metric consider a small 
vicinity of the horizon surface $\Sigma$.
For $\beta \neq \beta_H$ the metric there reads
\begin{equation}
ds^2_{E_\alpha}=ds^2_\Sigma+\hat{\rho}^2_+ ds^2_{C_\alpha}
\label{15}
\end{equation}
Replacing the cone metric $ds^2_{C_\alpha}$ by $ds^2_{C_{\alpha , b}}$ (\ref{13})
we obtain a sequence of regular metrics
\begin{equation}
ds^2_{E_{\alpha , b}}=ds^2_\Sigma+\hat{\rho}^2_+ ds^2_{C_{\alpha , b}}
\label{16}
\end{equation}
To calculate the curvature we define the (anholonomic) basis
of one-forms $\{ e^a,~a=1,...,4 \}$ orthonormal with respect to metric (\ref{16}):
\begin{eqnarray}
&&e^1=b \hat{\rho}_+ f^{1/2}(y)dy \nonumber \\
&&e^2=by {(m^2+\hat{a}^2+\hat{q}^2)^{1/2}\over \hat{\rho}_+}(d\tau-\hat{a}\sin^2\theta d \phi )
 \nonumber \\
&&e^3=\hat{\rho}_+d\theta \nonumber \\
&&e^4={(\hat{r}_+^2-\hat{a}^2)\over \hat{\rho}_+}\sin\theta (d\phi+{\hat{a}\over
(\hat{r}^2_+-\hat{a}^2)}d\tau )
\label{17}
\end{eqnarray}
where we changed variables so that
 $x=b y,~~f(y)={y^2+\alpha^2\over y^2+1}$.

 The Lorentz connection one-form $\omega^a_{\ b}=\omega^a_{\ b\ c}e^c$
is found from the equation:
\begin{equation}
de^a+\omega^a_{\ b}\wedge e^b=0
\label{18}
\end{equation}
We are interested in those components of the Lorentz connection which are
singular in the limit $b\rightarrow 0$. Analyzing the 
expressions $de^a$ for the basis (\ref{17})
we observe that only $de^2$ contains a singular term:
\begin{equation}
de^2={dy\over y}\wedge e^2+ ...=(by f^{1/2}(y))^{-1}e^1\wedge e^2+ ...~~,
\label{19}
\end{equation}
where ``$...$'' means terms finite in the limit $b\rightarrow 0$. It follows from
(\ref{19}) that the only singular component of the Lorentz connection is
\begin{equation}
\omega^2_{\ 1}=(by\hat{\rho}_+ f^{1/2}(y))^{-1} e^2+...
\label{20}
\end{equation}

The curvature two-form $R^a_{\ b}=R^a_{\ b\ c\ d}e^c\wedge e^d$ 
is defined as follows
$$
R^a_{\ b}=d\omega^a_{\ b}+\omega^a_{\ c}\wedge\omega^c_{\ b}~~.
$$
Again, the only singular component of $R^a_{\ b}$ is
$$
R^2_{\ 1}=d\omega^2_{\ 1}+ ...= {1\over 2yb^2\hat{\rho}^2_+}{f'_y \over f^2}e^2\wedge e^1 + ...
$$
and so in terms of curvature components the only singular component is
\begin{equation}
R_{2121}= {1\over 2yb^2\hat{\rho}^2_+}{f'_y \over f^2} + ...
\label{21}
\end{equation}
Introducing a pair of vectors (see Appendix Eqs.(\ref{B1}), (\ref{B2}))
 $n_a=n_a^\mu\partial_\mu,~a=1,2$ orthogonal to the horizon $\Sigma$
and dual to the one-forms $e^a,~a=1,2$
we may re-write this 
$R_{2121}={1\over 2}R_{\mu\nu\alpha\beta}n^{\mu}_an^\nu_bn^\alpha_an^\beta_b$.

In order to show that the component $R_{2121}$ 
behaves as a $\delta$-function in the limit $b\rightarrow 0$, 
let us consider the integral
\begin{equation}
I_{D_\epsilon}=\int_{D_\epsilon} R_{2121}~ v(x,\theta ,\psi )e^1\wedge e^2\wedge e^3\wedge e^4
\label{22}
\end{equation}
over a small disk $D_\epsilon$ surrounding the horizon surface 
$\Sigma$, $0 \leq x \leq \epsilon$.
In (\ref{22}) $v(x,\theta ,\psi )$ is a test function which is constant along the
trajectory of vector $K$ ($K[v]=0$). It can be expanded as
$$
v(x,\theta ,\psi )=v_0(\theta , \psi )+v_1(\theta , \psi )x^2 +...
$$
$$
=v_0(\theta , \psi )+b^2v_1(\theta , \psi )y^2 +...
$$
Recall that $(\theta , \psi )$ are the coordinates on the horizon $\Sigma$.
Substituting (\ref{17}), (\ref{21}) into (\ref{22}) we obtain
\begin{eqnarray}
&&I_{D_\epsilon}=\int_0^{\epsilon\over b}dy{f'_y\over 2 f^{3/2}}\sqrt{m^2+\hat{a}^2+\hat{q}^2}
\oint (d\tau-\hat{a}\sin^2\theta d \phi ) \nonumber \\
&&\int_\Sigma{1\over \hat{\rho}^2_+}(v_0+v_1 b^2 y^2+...)e^3\wedge e^4
\label{23}
\end{eqnarray}
In (\ref{23}) we first integrate over $e^1\wedge e^2$ in the subspace orthogonal to
$\Sigma$ under fixed $(\theta , \psi )$ and then take the integral 
$\int_\Sigma e^3\wedge e^4$ over the horizon.
This yields
\begin{equation}
\oint (d\tau-\hat{a}\sin^2\theta d \phi )={2\pi\beta \hat{\rho}^2_+ \over \hat{r}^2_+-\hat{a}^2_+}~~,
\label{24}
\end{equation}
where the integration is taken over the closed integral trajectory 
of the Killing vector $K$
under fixed $(\theta,~\psi)$.

In the limit $b\rightarrow 0$, we have 
${\epsilon\over b} \rightarrow \infty$ in the $y$-integration in
(\ref{23}) and so obtain
\begin{equation}
\int_0^\infty dy {f'_y\over 2f^{3/2}}=-f^{-1/2}(y)|^\infty_0={1-\alpha\over \alpha}
\label{25}
\end{equation}
Taking into account that $\beta_H={\hat{r}^2_+-\hat{a}^2 \over \sqrt{m^2+\hat{a}^2+\hat{q}^2}}$
from (\ref{23})-(\ref{25}) we finally obtain in the limit $b\rightarrow 0$:
\begin{equation}
I_{D_\epsilon}=2\pi(1-\alpha )\int_\Sigma v_0 (\theta , \psi ) e^3\wedge e^4
\label{26}
\end{equation}
Since this holds for arbitrarily small $\epsilon$ we conclude that 
in the limit  $b\rightarrow 0$ the quantity $R_{2121}$ behaves
as a $\delta$-function having support at the surface $\Sigma$.
Noting that the vectors $n_a,~a=1,2$ introduced above are normal 
to $\Sigma$ we may write 
\begin{eqnarray}
&&R^{\mu\nu}_{\ \ \alpha\beta} = \bar{R}^{\mu\nu}_{\ \
\alpha\beta}+
2\pi (1-\alpha) \left( (n^\mu n_\alpha)(n^\nu n_\beta)- (n^\mu
n_\beta)
(n^\nu n_\alpha) \right) \delta_\Sigma \nonumber \\
&&R^{\mu}_{ \ \nu} = \bar{R}^{\mu}_{ \ \nu}+2\pi(1-\alpha)(n^\mu
n_\nu)
\delta_\Sigma \nonumber                            \\
&&R = \bar{R}+4\pi(1-\alpha) \delta_\Sigma
\label{27}
\end{eqnarray}
where $\delta_\Sigma$ is the delta-function
$\int_{\cal M}^{}f\delta_\Sigma e^1\wedge e^2\wedge e^3\wedge e^4=
\int_{\Sigma}^{}f e^3\wedge e^4$ and we denote
$(n_\mu n_\nu)=\sum_{a=1}^{2}n^a_\mu
n^a_\nu$. In particular, it follows from (\ref{27}) that
\begin{equation}
\int_{E_{\alpha}}Re^1\wedge e^2\wedge e^3\wedge e^4 =\alpha\int_{E} \bar{R}e^1\wedge e^2\wedge e^3\wedge e^4
+4\pi(1-\alpha) A_\Sigma~~~,
\label{28}
\end{equation}
where $A_\Sigma=\int_\Sigma e^3\wedge e^4$ is area of $\Sigma$. 
For the particular case of the Kerr-Newman metric, $\bar{R}=0$.
Remarkably, expressions (\ref{27}), (\ref{28}) are exactly the same as that
obtained in \cite{FS} for a static metric.

For a variety of applications it is necessary to know the integrals of
quadratic combinations of curvatures over the space $E_\alpha$ with a
conical singularity at the surface $\Sigma$. According to (\ref{27})
 curvature $\cal R$ contains $(1-\alpha )\delta_\Sigma$-contribution as well as a
regular part $\bar{\cal R}$.
 Therefore, one can expect the result which can be symbolically written
as follows
\begin{equation}
\int_{E_\alpha} {\cal R}^2=\alpha \int_{E_{\alpha=1}}  \bar{\cal R}^2
+2(1-\alpha ) \int_\Sigma \bar{\cal R}_n+ O((1-\alpha )^2)~~,
\label{29}
\end{equation}
where $\bar{\cal R}_n$ means projection of the tensor $\bar{\cal R}$ onto the subspace
orthogonal to the singular surface $\Sigma$. The expression (\ref{29})
is ill-defined since ${\cal R}^2$ contains  term $((1-\alpha )\delta_\Sigma)^2$.
However, it is of higher order with respect to $(1-\alpha )$ and so
can be collected in the last term in (\ref{29}). 

The form (\ref{29}) was obtained in 
\cite{FS1} for a static metric. To verify this for the Kerr-Newman case 
we must write the 
metric (\ref{5}) near the horizon $\Sigma$, including all terms of 
order $x^2$. Taking into account the regularization 
function $f(x,b)$ as above the metric reads:
\begin{eqnarray}
&&ds^2_{E_{\alpha , b}}=by f(y)dy^2+{\gamma^2 b^2 y^2\over 4\hat{\rho}^2_+}
(d\tau-\hat{a}\sin^2\theta d\phi )^2+(\hat{\rho}^2_++{\gamma \over 2}b^2y^2\hat{r}_+ )
d\theta^2 \nonumber \\
&&+\left( {(\hat{r}^2_+-\hat{a}^2)^2\over \hat{\rho}^2_+}+
{(\hat{r}^2_+-\hat{a}^2)\over \hat{\rho}^2_+}(1-
{(\hat{r}^2_+-\hat{a}^2)\over 2\hat{\rho}^2_+})\gamma \hat{r}_+ b^2 y^2 \right)
\sin^2\theta (d\phi+{\hat{a}\over \hat{r}^2_+-\hat{a}^2} d\tau )^2 \nonumber \\
&&-{\gamma \hat{a}b^2y^2\over 2\hat{\rho}^2_+}\sin^2\theta (d\phi+
{\hat{a}\over \hat{r}^2_+-\hat{a}^2} d\tau )d\tau
\label{30}
\end{eqnarray}
The general structure of quadratic combinations of curvature terms 
(denoted by  ${\cal R}^2$) for the metric (\ref{30}) symbolically is 
\begin{equation}
{\cal R}^2=b^2 A+{f'_y \over b^2} B+{(f'_y)^2\over b^4} C+O(b^4)
\label{31}
\end{equation}
where $A,B,C$ are some functions that are independent of $b$ and do not 
contain derivatives of the regularization function $f(y)$.

Since the measure of integration in the region near $\Sigma$ is 
proportional to $b^2$ we conclude that second and third terms 
in (\ref{31}) after integration produce
in the limit $b\rightarrow 0$ the respective second and third terms 
in (\ref{29}). In order to get this we use the fact that
the derivatives of $f(y)$ behave as $f'(y)\sim (1-\alpha)$.

After straightforward but tedious calculations we obtain in the 
limit $b\rightarrow 0$:
\begin{equation}
\int_{E_{\alpha}}R^{\mu\nu}R_{\mu\nu}=\alpha\int_{ E}
\bar{R}^{\mu\nu}\bar{R}_{\mu\nu}
+4\pi(1-\alpha)\int_{\Sigma}\bar{R}_{aa}+O((1-\alpha)^2)~~~,
\label{A3}
\end{equation}
\begin{equation}
\int_{E_{\alpha}}R^{\mu\nu\lambda\rho}R_{\mu\nu\lambda\rho}
=\alpha\int_{E}
\bar{R}^{\mu\nu\lambda\rho}\bar{R}_{\mu\nu\lambda\rho}
+8\pi(1-\alpha)\int_{\Sigma} \bar{R}_{abab}
+O((1-\alpha)^2)~~~,
\label{A4}
\end{equation}
where $\bar{R}_{aa}=\sum_{a=1,2}\bar{R}_{\mu\nu}n_a^{\mu}n_a^{\nu}$ and
$\bar{R}_{abab}=\sum_{a,b=1,2}\bar{R}_{\mu\nu\lambda\rho}n^{\mu}_an^{\lambda}_
an^{\nu}_b n^{\rho}_b $.

In obtaining (\ref{A3}), (\ref{A4}) we made use of the fact 
(as in (\ref{23})) that near $\Sigma$ the measure of integration $\mu_{E_{\alpha , b}}$ takes the form (see (\ref{5}))
$\mu_{E_{\alpha , b}}=\hat{\rho}^2_+ \mu_\Sigma \mu_{C_{\alpha , b}}$, where
$\mu_\Sigma=(\hat{r}^2_+-\hat{a}^2)\sin \theta d\theta d\psi$ ($0\leq \psi\leq 2\pi$)
is the measure on $\Sigma$ and $\mu_{C_{\alpha , b}}=\alpha b^2 f^{1/2}(y)dy d\bar{\chi}$
($0\leq \bar{\chi}\leq 2\pi$) is the 
measure on the regularised cone $C_{\alpha , b}$.
For the integral of $R^2$ we obtain
$$
\int_{E_\alpha}R^2=O((1-\alpha)^2)
$$
in agreement with the expected formula
\begin{equation}
\int_{E_{\alpha}}R^2=\alpha\int_{E} \bar{R}^2
+8\pi(1-\alpha)\int_{\Sigma}\bar{R}+O((1-\alpha)^2)~~~,
\label{A2}
\end{equation}
since the Kerr-Newman metric satisfies $\bar{R}=0$.

Again we obtain for a stationary metric with a conical singularity 
the same expressions (\ref{A3})-(\ref{A2}) as for the  
static case \cite{FS1}.

For the Kerr-Newman metric we have on the horizon $\Sigma$:
\begin{eqnarray}
&&{1\over 2}\bar{R}_{abab}=\bar{R}_{2121}={\hat{r}^2_+(4\hat{q}^2+8m\hat{r}_+)-(\hat{q}^2+6m\hat{r}_+)\hat{\rho}^2
\over \hat{\rho}_+^6} \nonumber \\
&&{1\over 2}\bar{R}_{aa}=\bar{R}_{11}=\bar{R}_{22}={\hat{q}^2\over \hat{\rho}^4_+}
\label{32}
\end{eqnarray}
and after integration over $\Sigma$ we get
\begin{eqnarray}
&&\int_\Sigma \bar{R}_{abab}=8\pi{(\hat{r}^2_++\hat{q}^2)\over \hat{r}^2_+}+
4\pi{\hat{q}^2\over \hat{r}^2_+}{(\hat{r}^2_+-\hat{a}^2)\over \hat{a}\hat{r}_+}
\ln ({\hat{r}_++\hat{a}\over \hat{r}_+-\hat{a}}) \nonumber \\
&&\int_\Sigma \bar{R}_{aa}=4\pi{\hat{q}^2\over\hat{r}^2_+}\left( 1 +
{(\hat{r}^2_+-\hat{a}^2)\over 2\hat{a}\hat{r}_+}
\ln ({\hat{r}_++\hat{a}\over \hat{r}_+-\hat{a}})\right)
\label{33}
\end{eqnarray}
The analytic continuation of these expressions back to real values
of the parameters $a$ and $q$ requires  the substitution
\begin{eqnarray}
&&\hat{q}^2=-q^2,~~\hat{a}^2=-a^2,~~\hat{r}_+=r_+ \nonumber \\
&&{1\over \hat{a}}\ln ({\hat{r}_++\hat{a}\over \hat{r}_+-\hat{a}})=
{2\over a} \tan^{-1}({a\over r_+})
\label{34} 
\end{eqnarray}

\bigskip

\section{Heat kernel expansion and entropy}
\setcounter{equation}0

In the Euclidean path integral approach to a statistical field system 
at temperature $T=(2\pi\beta )^{-1}$ one considers the fields which are 
periodic with
respect to imaginary time $\tau$ with period $2\pi\beta$. This works well
for a static black hole when the metric does not depend on 
the time coordinate $\tau$
\cite{GH}.
One then closes the integral curves of the Killing vector 
$\partial_\tau$ with the period $2\pi\beta$. 

For a rotating black hole 
metric we need to close the integral curves of the vector $K$ (\ref{4}). 
The result of this is that for arbitrary $\beta$ we obtain the conical 
space $E_\alpha$, the geometry of which was described in the previous 
section.  The partition function then reads
\begin{equation}
Z(\beta )=\int [{\cal D}\varphi ] \exp[-I_E(\varphi , g_{\mu\nu})] ~~,
\label{35}
\end{equation}
where the matter Euclidean action $I_E$ is considered on 
the space $E_\alpha$ with
appropriate boundary ({\it i.e.} periodicity) conditions imposed on the
matter field(s) $\varphi$. The contribution to the entropy
is 
\begin{equation}
S=-(\beta \partial_\beta-1)\ln Z(\beta )|_{\beta=\beta_H}
\label{36}
\end{equation}

Although the Kerr-Newman metric is a solution of the Einstein equations with
the matter source in form of a Maxwell field, the gravitational action is always
modified by higher-order curvature terms due to quantum corrections. 
Such  $R^2$-terms must be added to the action at the outset with some 
bare constants ($c_{1,B}, c_ {2,B}, c_{3,B}$) (tree-level) 
to absorb  the one-loop infinities. 
The bare (tree-level) gravitational action functional thus takes the form
\begin{equation}
W_{gr}=\int{}\sqrt{g}d^4x \left( -{1 \over 16\pi G_B} R
+c_{1,B}R^2+c_{2,B}
R_{\mu \nu}R^{\mu \nu} 
+c_{3,B} R_{\mu\nu\alpha\beta} R^{\mu\nu\alpha\beta} \right)
\label{37}
\end{equation}
Of course, we assume in addition to (\ref{37}) a
classical matter action which can in principle be rather complicated.
The corresponding tree-level entropy can be obtained 
 as a replica of the action
(\ref{37}) after introducing the regulated conical singularity and
applying formulas (\ref{35})-(\ref{36}). 
Using formulas (\ref{A3})-(\ref{A2}) of the previous Section we obtain
for the tree-level entropy:
\begin{equation}
S(G_B, c_{i,B})= {1 \over 4G_B} A_\Sigma-\int_{\Sigma} \left( 8\pi
c_{1,B} \bar{R}
+4\pi c_{2,B}\bar{R}_{aa}+8\pi  c_{3,B}\bar{R}_{abab} \right)
\label{38}
\end{equation}
where $\bar{R}_{aa}=\sum_{a=1,2}\bar{R}_{\mu\nu}n_a^{\mu}n_a^{\nu}$ and
$\bar{R}_{abab}=\sum_{a,b=1,2}\bar{R}_{\mu\nu\lambda\rho}n^{\mu}_an^{\lambda}_
an^{\nu}_b n^{\rho}_b $, $\{n_a^\mu,~a=1,2 \}$ are vectors normal to $\Sigma$.
This is exactly the same expression that we had for the static case
\cite{Wald}, \cite{FS}. Expression (\ref{38}) is really valid off-shell,
as we do not require the metric to satisfy any equations of motion.
On-shell we must substitute in (\ref{38}) the field
equation $\bar{R}=0$ satisfied by the Kerr-Newman metric.

At the one-loop level we consider the matter action in the form
$$
I_{E}={1\over 2}\int_{E_\alpha}(\nabla \varphi )^2
$$
and get for the partition function
$$
\ln Z (\beta )=-{1\over 2}\ln det (-\Box_{E_\alpha})
$$
expressed via the determinant of the Laplacian $\Box=\nabla_\mu\nabla^\mu$
over the conical space $E_\alpha$.
In the De Witt-Schwinger proper time representation we have for the
logarithm of the determinant:
\begin{equation}
\ln det (- \Box )=-\int^\infty_{\epsilon^2} {ds\over s} Tr (e^{s\Box})
\label{39}
\end{equation}
In four dimensions we have the asymptotic expansion
\begin{equation}
Tr (e^{s\Box}) ={1\over (4\pi s)^2}\sum^{\infty}_{n=0} a_n s^n
\label{40}
\end{equation}
and for the divergent part of $(\ln Z)_{div}$
we get
\begin{equation}
(\ln Z)_{div}={1\over 32\pi^2}({1\over 2}a_0 \epsilon^{-4}+a_1\epsilon^{-2}+2a_2 \ln{L\over \epsilon})~~,
\label{41}
\end{equation}
where $L$ is an infra-red cut-off.
It is known that for a manifold with conical singularities the heat kernel 
coefficients in (\ref{40}) are really a sum
\begin{equation}
a_n=a^{st}_n + a_{n, \alpha} 
\label{42}
\end{equation}
of standard plus conical coefficients.
The standard coefficients  $\bar{a}^{st}_n$ are the same as for for a
smooth manifold $E$  \cite{BD}:
\begin{eqnarray}
&&a_0^{st}=\int_{E_\alpha} 1 \ \ , \ \ a^{st}_1={1 \over 6}\int_{E_\alpha} \bar{R}
\nonumber \\
&&a^{st}_2 =\int_{E_\alpha}\left( {1 \over 180} 
\bar{R}_{\mu\nu\alpha\beta}\bar{R}^{\mu\nu\alpha\beta} -
{1 \over 180} \bar{R}_{\mu\nu} \bar{R}^{\mu\nu} -{1 \over 30} \Box
\bar{R}
+{1 \over 72}\bar{R}^2 \right)
\label{43}
\end{eqnarray}
whereas the  parts coming from the singular surface $\Sigma$ (stationary
point of the isometry) are
\begin{eqnarray}
&&a_{0, \alpha}=0; \ \ \ a_{1, \alpha }={\pi \over
3}{(1-\alpha^2)
\over
\alpha}
\int_{\Sigma}^{}\sqrt{\gamma} d^2 \theta \ ; \nonumber \\
&&a_{2, \alpha }={\pi \over 3} {(1-\alpha^2) \over \alpha}
\int_{\Sigma}^{}({1 \over 6} \bar{R}+\lambda_1 (\kappa^a \kappa^a-2 tr(\kappa . \kappa ))) \sqrt{\gamma} d^2 \theta
\nonumber
\\
&&-{\pi \over 180}
{(1-\alpha^4) \over \alpha^3}
\int_{\Sigma}^{}(\bar{R}_{aa}
-2\bar{R}_{abab}  +{1\over 2}\kappa^a  \kappa^a +\lambda_2
 (\kappa^a \kappa^a-2 tr(\kappa . \kappa )) )\sqrt{\gamma} d^2 \theta
\label{44}
\end{eqnarray}
where $\lambda_{1,2}$ are some constants and
 $\kappa^a_{\mu\nu},~a=1,2$ is the extrinsic curvature of the surface $\Sigma$
with respect to normal vector $n_a,~a=1,2$; $\kappa^a=g^{\mu\nu}\kappa^a_{\mu\nu}$, $tr(\kappa .
\kappa )=\sum_{a=1,2}\kappa^a_{\mu\nu} \kappa^{\mu\nu}_a$.

The expression (\ref{44}) for some special spaces has been known
for some time \cite{15}. For a general static metric it was derived recently by
Fursaev \cite{F1}, in the case that all extrinsic curvatures $\kappa^a_{\mu\nu}$ 
vanished. Dowker \cite{Dowker} derived the heat kernel coefficients in 
the form (\ref{44}) for an arbitrary conical metric of a general type with a
surface $\Sigma$ having non-trivial extrinsic geometry.
Very general arguments were used in \cite{Dowker} to derive the general structure of
(\ref{44}): $O(2)$-invariance, dimensionality and conformal invariance. The result
(\ref{44}) contains some unknown constants $\lambda_1$ and $\lambda_2$ in front
of the conformal-invariant combination $(\kappa^a \kappa^a-2 tr(\kappa . \kappa ))$.
The analysis of \cite{Dowker} does not provide a prescription for obtaining 
the explicit values  for these constants.

Applying the formula (\ref{36}) to (\ref{41}) and  taking into account
that the standard coefficients $a^{st}_n\sim\alpha$
 we obtain for the divergent quantum correction
to the entropy
\begin{eqnarray}
&&S_{div}= {1 \over 48\pi \epsilon^2}  A_\Sigma + (
{1
\over 144 \pi}
 \int_\Sigma \bar{R}
-{1 \over 16 \pi } {1 \over 45} \int_{\Sigma}
(\bar{R}_{aa}-2\bar{R}_{abab} ) \nonumber \\
&&-{1\over 16\pi}{1\over 90}\int_\Sigma 
\kappa^a \kappa^a +{1\over 24\pi}(\lambda_1-{\lambda_2\over 30})\int_\Sigma
(\kappa^a \kappa^a-2 tr(\kappa . \kappa )))
\ln {L \over \epsilon}~~.
\label{45}
\end{eqnarray}
We see that the divergent part of the entropy (\ref{45}) depends both on 
the projections of the curvatures, $\bar{R}_{aa}$ and $\bar{R}_{abab}$, onto the
subspace normal to the horizon surface $\Sigma$ and on the quadratic combinations 
of the extrinsic curvatures of $\Sigma$. For the static case all extrinsic curvatures vanish 
and (\ref{45}) repeats the form of the tree-level entropy (\ref{38}). This 
fact allows one to prove \cite{FS} for arbitrary static black holes 
the statement \cite{9}-\cite{SS}
that all the UV divergences of entropy are absorbed in the standard 
renormalization of the gravitational couplings ($G,c_i)$ in the tree-level 
gravitational action (\ref{37}). Applying the same line of reasoning to 
the Kerr-Newman black hole entails studying the external geometry 
of the horizon $\Sigma$
of the charged rotating black hole. For this case we find that
$\sum_{a=1,2}\kappa^a \kappa^a =
tr (\kappa.\kappa )=0$ (see Appendix). This makes the coefficients 
(\ref{44}) and the expression for $S_{div}$ (\ref{45}) 
for the Kerr-Newman metric the same as for a static metric. 

Consequently, $S_{div}$ in (\ref{45})
 repeats the form of the tree-level 
entropy and the {\it renormalization } statement is valid for a stationary 
hole as well. In one sense this is not surprising since the classical 
thermodynamics of static and stationary holes are formulated in the same 
way. One could therefore expect this to also be valid in the quantum case.

Consider (\ref{45}) on the Kerr-Newman background.
Substituting here (\ref{33}), the condition $\bar{R}=0$,
and making the analytic continuation (\ref{34}), we finally obtain for 
the quantum entropy of the Kerr-Newman black hole:
\begin{equation}
S_{div}={1 \over 48\pi \epsilon^2}  A_\Sigma 
+{1\over 45}
\left( 1 -{3q^2\over 4r^2_+}(1+{(r^2_++a^2)\over a r_+} \tan^{-1}({a\over r_+}))
\right) \ln {L \over \epsilon}
\label{46}
\end{equation}
where $A_\Sigma=4\pi (r^2_++a^2)$ is area of the horizon $\Sigma$.
In the limit $a\rightarrow 0$ this expression reduces to that of the
Reissner-Nordstrom black hole obtained in \cite{SS1} using the conical
method and in \cite{14} within the framework of a 
statistical-mechanical calculation in spirit of t'Hooft's approach.
Surprisingly, in the uncharged case ($q=0$) the second term in the expression
(\ref{46}) does not depend on the rotation parameter $a$ and 
it is the same as for
the Schwarzchild black hole \cite{SS}. We do not have 
an explanation of this fact.

\section{Conclusions}

The Euclidean approach to black hole thermodynamics implying the conical singularity
 method is known to be very useful in the static case. It allows one to 
get both the classical and quantum thermodynamic quantities of static 
black holes.
We have proposed that the thermodynamics of the classical static and 
stationary black holes are formulated in a similar way. The underlying
assumption is that the conical singularity technique
can be applied to the rotating hole as well. 

In this paper we logically
followed this line of reasoning. We studied the Euclidean geometry of the 
Kerr-Newman metric for an arbitrary period along the time-like Killing vector 
generating the abelian isometry of the space. The conical geometry of the 
space near the horizon
was established and the $\delta$-function like behavior of the curvatures 
obtained. This behavior strongly resembles that of a static black hole.

The essential point of formulating the quantum thermodynamics of static black 
hole is the proving the statement that all the UV-divergences
of the entropy of black hole due to quantum matter are removed by
the standard renormalization of couplings in the tree-level gravitational
action. This allows one to consider the entropy as well-defined quantum
field theoretical quantity.
We demonstrate for the Kerr-Newman black hole that $S_{div}$ being expressed via
geometrical invariants repeats the form of the tree-level entropy
in the same way as for a static case. This proves that the {\it renormalization }
statement works universaly both for the static and stationary holes providing
the correct treatment of the quantum thermodynamics.

However, it is still an open question as to
what degrees of freedom are counted by the entropy 
of black hole. A useful approach to this problem is to compare our 
result with the  statistical-mechanical
calculation of the quantum entropy of Kerr-Newman black hole along the lines 
of \cite{6}-\cite{14}. For a charged non-rotating black hole it is known 
that there is perfect agreement between these two methods 
(see \cite{14} and \cite{SS1}). Checking
this for stationary case\footnote{The recent statistical calculation 
performed in \cite{K} appears to be unsatisfactory since it relates the 
entropy of rotating hole with data on the ergosphere rather than on the 
horizon.} should provide us with a better understanding of the 
relationship between the different entropies assigned to a black hole \cite{SSS}.

\section*{Acknowledgements}
This work was supported by the Natural Sciences and Engineering Research
Council of Canada and by a NATO Science Fellowship.

{\appendix \noindent{\large \bf Appendix: Extrinsic geometry of horizon}}\\
\def\theequation{A.\arabic{equation}}
\setcounter{equation}0
 
With respect to the  Euclidean metric (\ref{5}) we may define a pair of 
orthonormal vectors $\{n_a=n_a^\mu \partial_\mu~,~~a=1,2 \}$:
\begin{equation}
n_1^r=\sqrt{\hat{\Delta} \over \hat{\rho}^2}
\label{B1}
\end{equation}
\begin{equation}
n_2^\tau={(r^2-\hat{a}^2)\over \sqrt{\hat{\Delta} \hat{\rho}^2}}~,~~n_2^\phi={-
\hat{a}\over  \sqrt{\hat{\Delta} \hat{\rho}^2}}
\label{B2}
\end{equation}
Covariantly these are
\begin{equation}
 n^1_r=\sqrt{\hat{\rho}^2 \over \hat{\Delta}}
\label{B3}
\end{equation}   
\begin{equation}
n^2_\tau= \sqrt{\hat{\Delta} \over \hat{\rho}^2}~,~~n^2_\phi=-\sqrt{\hat{\Delta} \over \hat{\rho}^2} \hat{a}\sin^2\theta
\label{B4}
\end{equation} 
The vectors $n^1$ and $n^2$ are normal to the horizon surface $\Sigma$ 
(defined as $r=r_+,~\Delta (r=r_+)=0)$, which is a
two-dimensional surface with induced metric $\gamma_{\mu\nu}=g_{\mu\nu}-
n^1_\mu n^1_\nu -n^2_\mu n^2_\nu$. The (non-zero) 
components of the induced metric are
\begin{eqnarray}
&&\gamma_{\theta\theta}=\rho^2, ~~\gamma_{\tau\tau}={\hat{a}^2\sin^2\theta \over 
\hat{\rho}^2 } \nonumber \\
&&\gamma_{\tau\phi}={\hat{a} (r^2-\hat{a}^2)\sin^2\theta \over \hat{\rho}^2} \nonumber \\
&&\gamma_{\phi\phi}={(r^2-\hat{a}^2)^2\sin^2\theta \over \hat{\rho}^2}
\label{B5}
\end{eqnarray}
With respect to the normal vectors $n^a, ~a=1,2$ 
we may define \cite{Eisen} the extrinsic 
curvatures of the surface $\Sigma$: 
$\kappa^a_{\mu\nu}=-\gamma_\mu^\alpha \gamma_\nu^\beta \nabla_\alpha n^a_\beta$.
We find
\begin{eqnarray}
&&\kappa^1_{\theta\theta}=-r\sqrt{\hat{\Delta} \over \hat{\rho}^2 } \nonumber \\
&&\kappa^1_{\tau\tau}={r\hat{a}^2\sin^2\theta\over \hat{\rho}^4}\sqrt{\hat{\Delta}
\over \hat{\rho^2}} \nonumber \\
&&\kappa^1_{\tau\phi}=-{\hat{a}r(r^2-\hat{a}^2)\sin^2\theta \over \hat{\rho}^4} \sqrt{\hat{\Delta}\over \hat{\rho}^2} \nonumber \\
&&\kappa^1_{\phi\phi}=-{r(r^2-\hat{a}^2)^2\sin^2\theta  \over \hat{\rho}^4}
\sqrt{\hat{\Delta}\over \hat{\rho}^2} 
\label{B6}
\end{eqnarray}
and
\begin{eqnarray}
&&\kappa^2_{r\tau}=-{\hat{a}^2\sin \theta \cos \theta \over \hat{\rho}^2}
\sqrt{{\hat{\Delta}
\over \hat{\rho}^2}} \nonumber \\
&&\kappa^2_{r\phi}=-{\hat{a}(r^2-\hat{a}^2)\sin \theta \cos \theta \over 
\hat{\rho}^2}\sqrt{{
\hat{\Delta}\over \hat{\rho}^2}}
\label{B7}
\end{eqnarray}
For the trace of the extrinsic curvatures, $\kappa^a=\kappa^a_{\mu\nu}g^{\mu\nu}$,
we obtain:
\begin{equation}
\kappa^1=-{2r\over \hat{\rho}^2}\sqrt{\hat{\Delta}\over \hat{\rho}^2}~,~~\kappa^2=0
\label{B8}
\end{equation}
which clearly vanishes when restricted to the surface 
$\Sigma$ $(\Delta(r=\hat{r}_+)=0)$.

The quadratic combinations 
\begin{eqnarray}
&&\kappa^1_{\mu\nu} \kappa_{1}^{\mu\nu}= {2r^2\hat{\Delta}\over \hat{\rho}^6}
\nonumber \\
&&\kappa^2_{\mu\nu} \kappa_{2}^{\mu\nu}= {2\hat{a}^2\cos^2\theta \hat{\Delta}
 \over \hat{\rho}^6} 
\label{B9}
\end{eqnarray}
vanish $\Sigma$ separately both in the static  ($a=0$) and stationary 
 $(a\neq 0)$ cases. Consequently, we have
$tr (\kappa . \kappa )=\kappa^a_{\mu\nu} \kappa^{a\mu\nu}=0$ on the horizon.

\end{document}